\documentclass[a4paper]{article}

\usepackage[english]{babel}
\usepackage[utf8]{inputenc}
\usepackage[round]{natbib}
\usepackage{authblk,bbm,bm}
\usepackage{amsmath,setspace,geometry,lineno,amsfonts,amssymb,soul}
\usepackage{graphicx}
\usepackage{multirow}
\usepackage[colorinlistoftodos]{todonotes}
\usepackage{fullpage}
\usepackage{lineno}
\usepackage{float}
\usepackage{caption}	
\usepackage{subcaption}
\usepackage{gensymb}
\usepackage{changepage}
\usepackage{hyperref}
\usepackage{url}

\begin{document}
\begin{spacing}{1.1}


\centerline{\LARGE Evidence for a postreproductive phase in female false killer}
\vspace{6pt}
\centerline{\LARGE whales \em{Pseudorca crassidens}}

\vspace{12pt}

\centerline{{\bf Theoni Photopoulou}$^{\rm 1, 2, *}$, {\bf In\^es M Ferreira}$^{\rm 3}$, {\bf Peter B Best}$^{\rm 3, \dagger}$, 
\bf{Toshio Kasuya}$^{\rm 4}$ and {\bf Helene Marsh}$^{\rm 5}$}

\vspace{12pt}

\centerline{$^{\rm 1}$Department of Zoology, Nelson Mandela Metropolitan University}
\centerline{Port Elizabeth 6031, South Africa}
\centerline{$^{\rm 2}$Centre for Statistics in Ecology, Environment and Conservation}
\centerline{Department of Statistical Sciences, University of Cape Town}
\centerline{Rondebosch 7701, South Africa}
\centerline{$^{\rm *}$Corresponding author: theoni.photopoulou@gmail.com}

\vspace{12pt}

\centerline{$^{\rm 3}$Mammal Research Institute, University of Pretoria}
\centerline{c/o Iziko South African Museum, Box 61}
\centerline{Cape Town 8000, South Africa}

\centerline{$^{\rm \dagger}$ Deceased}

\vspace{12pt}

\centerline{$^{\rm 4}$35-30-32-3 Nagayama, Tama, Tokyo 206-0025, Japan}

\vspace{12pt}

\centerline{$^{\rm 5}$James Cook University, JCU Post Office}
\centerline{Townsville 4811, Queensland, Australia}

\vspace{12pt}

\begin{abstract} 
\textbf{Introduction} 
A substantial period of life after reproduction ends, known as postreproductive lifespan (PRLS), is at odds with classical life history theory and its causes and mechanisms have puzzled evolutionary biologists for decades. Prolonged PRLS has been confirmed in only two non-human mammals, both odontocete cetaceans in the family Delphinidae. We investigate the evidence for PRLS in a third species, the false killer whale, \textit{Pseudorca crassidens}, using a quantitative measure of PRLS and morphological evidence from reproductive tissue.  

\textbf{Results} 
We examined specimens from false killer whales from combined strandings (South Africa, 1981) and harvest (Japan 1979-80) and found morphological evidence of changes in the activity of the ovaries in relation to age. Ovulation had ceased in 50\% of whales over 45 years, and all whales over 55 years old had ovaries classified as postreproductive. We also calculated a measure of PRLS, known as postreproductive representation (PrR) as an indication of the effect of inter-population demographic variability. PrR for the combined sample was 0.14, whereas the mean of the simulated distribution for PrR under the null hypothesis of no PRLS was 0.02. The 99th percentile of the simulated distribution was 0.08 and no simulated value exceeded 0.13. These results suggest that PrR was convincingly different from the measures simulated under the null hypothesis. The evidence supporting PRLS strengthened when the data from South Africa were considered on their own (PrR = 0.37), and remained effectively unchanged when data from Japan were considered on their own (PrR = 0.12).

\textbf{Conclusions} 
We found morphological and statistical evidence for PRLS in South African and Japanese pods of false killer whales, suggesting that this species is the third non-human mammal in which this phenomenon has been demonstrated in wild populations. Our estimates for PrR in false killer whales (0.12-0.37) spanned the single values available for the short-finned pilot whale (0.28) and the killer whale (0.22) and are comparable to estimates for historical or hunter-gather human populations (0.3-0.47). 

\end{abstract}

{\bf Keywords:} False killer whale, postreproductive lifespan, reproduction, odontocete

\vspace{0.5em}

\section*{Background}

Animals are said to have a postreproductive lifespan (PRLS) when reproductive senescence occurs faster than general somatic senescence \citep{alb13, cro15}. The onset of PRLS is defined as the irreversible loss of physiological capacity to produce offspring due to intrinsic biological factors \citep{coh04}. An extended PRLS is at odds with classical life history theory, and its causes and mechanisms have puzzled evolutionary biologists for decades \citep{wil57}, leading to lengthy debate. 

Despite most of the research being focused on humans (e.g., \citealp{haw98, lah04, lah14}), a detectable PRLS has been
reported in a diversity of taxa. For example, a comparison by \cite{jon14} of standardized patterns of fecundity with age including 11 mammals, 12 other vertebrates, 10 invertebrates, 12 vascular plants and a green alga found that in addition to humans, PRLS occurs in killer whales \textit{Orcinus orca}, bdelloid rotifers \textit{Macrotrachela sp.}, nematode worms \textit{Caenorhabditis elegans} and Bali mynah birds \textit{Leucopsar rothschildi}. These results were thought to support claims (e.g. \citealp{coh04}) that the phenomenon may be widespread. A contrasting perspective is that the apparently wide distribution of a significant PRLS is a methodological artefact \citep{lev11}, resulting from the practice of using postreproductive time (PrT) as a population measure of PRLS. PrT is measured in units of time between last parturition and death and therefore depends on the overall longevity of a species. Notably, \cite{lev11} found that PrT was inconsistently calculated across taxa, inherently biased, and leads to a high false-positive rate.  

In light of these shortcomings, an alternative measure of PRLS was introduced to capture the {\em proportion} of the adult lifespan that is postreproductive. This measure, postreproductive representation (PrR) is interpreted as the proportion of adult years lived that are postreproductive \citep{lev11}. The calculation of PrR is based on age-specific demographic rates from a life table and therefore allows for valid quantitative comparisons between populations with different demographic parameters. Importantly, the authors of the study also made available the procedures for comparing the point estimate for PrR to a null distribution representing the expected range of PrR under the assumption of no PRLS \citep{lev11}. 

Despite this methodological advance, there is no single accepted indicator of the cessation of female reproduction (e.g., \citealp{nic16}). A decrease in ovulation and pregnancy rates with age is typical of many mammals, few of which show a prolonged postreproductive phase. An age-related decline in fecundity is not just the result of changes in the ovaries, but is a consequence of the total sum of changes to the reproductive system. Thus, we consider that the existence of a postreproductive phase is best investigated on the basis of several lines of evidence.

The existence of a substantial female PRLS in killer whales was established by longitudinal observations of known individuals \citep{big82, ole90, war09, fos12, bre15}. Notwithstanding these studies of wild killer whales, investigating the existence of PRLS is generally difficult for pelagic marine mammals because they are notoriously difficult to observe. Data from both stranded animals and shore-drive fisheries are suitable for investigating the existence of PRLS in such species, as demonstrated for short-finned pilot whales, \textit{Globicephala macrorhynchus} \citep{kas84a, mar84} and long-finned pilot whales \textit{Globicephala melas} \citep{mar93}. 

In this paper, we use data from 91 carcasses from both stranded animals and shore-drive fisheries to examine the evidence for a substantial female PRLS in a third oceanic odontocete, the false killer whale, \textit{Pseudorca crassidens}. False killer whales share several life history characteristics with killer whales and short-finned pilot whales. All 3 species are highly social, have low life-time productivity, are sexually dimorphic, and are known or believed to exist in stable matrilineal groups of closely related females, with strong mother-offspring associations and a long period of dependency \citep{big82, kas84a, whi00}.

First, we examine the biological evidence for impaired female reproductive performance with age, based on the histology of reproductive tissues expected to change with age and reproductive cessation on the basis of the research on short-finned pilot whale life history \citep{kas84a, mar84}. We investigate age-related changes in ovarian weight per kg of body mass, ovarian activity, mammary gland thickness, and the occurrence of pregnancy. We then test whether a postulated postreproductive phase is supported statistically by calculating the postreproductive representation (PrR) of the females in our sample \citep{lev11}. Based on the results of our analyses, we conclude that it is likely that female false killer whales have a prolonged PRLS.

\subsection*{Study species}
False killer whales are distributed globally in tropical and warm temperate seas and are occasionally sighted in cold temperate regions \citep{bai02}. They are typically pelagic, although they also use the shallow waters around oceanic islands, such as the Hawaiian Islands \citep{bai08, bai16}. Their propensity to mass strand and the affiliative behaviour of stranded individuals have been taken as indicators of their extreme sociality \citep{bai02}. False killer whales usually travel in groups of 20 to 100 and long-term (at least 15 year) association patterns between individuals have been documented \citep{bai08, bai16}.


Life history information for false killer whales mostly comes from carcass analysis \citep{kas86, mar86, fer14, blo93, mar93}. Because the deposition rate of growth layer groups in the teeth of false killer whales has not been calibrated, there is some uncertainty in the estimates of these parameters but the uncertainty is thought to be low because the deposition rate has been confirmed in other odontocete cetaceans \citep{hoh09}. Female false killer whales mature at 8-14 years \citep{bai02, fer14}, though some males probably do not mature until later (up to 18.5 years, \citealp{fer14}). False killer whales are sexually dimorphic \citep{bai02, fer14} and the maximum longevity has been estimated at around 60 years for both sexes \citep{kas86, fer14}; females may live slightly longer than males. Calving rates are low and calving intervals long \citep{bai02, fer14}.

The social structure of false killer whales groups is not known and the dispersal patterns of male false killer whales from their natal school are poorly understood \citep{fer14}. Both stranded and shore-drive samples are characterised by few large juvenile and sub-adults \citep{kas86, alo99, fer14}. In contrast to short-finned pilot whales \citep{kas84a}, aggregations of maturing males have not been observed. 

\subsection*{Data collection}
Reproductive tissues are expected to degrade with age as females cease to be reproductively active. To test this hypothesis in false killer whales, we analysed samples from the ovaries and mammary glands following the techniques established for a parallel investigation into PRLS in short-finned pilot whales \citep{kas84a, mar84}. We investigated age-related trends in 1) ovarian weight per kg of body mass, 2) ovarian activity, 3) mammary gland thickness, and 4) the occurrence of pregnancy based on the presence of corpora lutea of pregnancy (CLP). Different numbers of samples were available for each tissue type, so sample sizes vary slight between analyses and are reported throughout.

We used an existing dataset, which was previously used to compare age and reproductive information from false killer whales stranded in South Africa in 1981, with similar material from animals examined from drive fisheries in Japan (1979-80) \citep{fer14}. The South African material was collected from 65 false killer whales that stranded en masse on the west coast of the Western Cape Province on 19 August, 1981. Of these, 56 were found over a 1.5 km stretch of beach in St Helena Bay (32.781$\degree$S 18.1$\degree$E). As scientists reached the site only two days after the stranding event was reported, the material was not fresh and fixation was suboptimal. Data were available from 41 (including 32 mature) females. The Japanese material (96 females, including 57 mature) originated from 6 schools of apparently healthy animals driven ashore at Iki Island (33.8$\degree$N 129.718$\degree$E) in February and March of 1979 and 1980. These drives were designed as culling operations to reduce fishery interactions \citep{kas85} and captured entire schools of whales. In each case, as many false killer whales as possible were randomly examined as soon as possible after death. Age data were available for 91 (including 32+57=89 mature) females, though not all samples were available from all individuals, resulting in differing sample sizes. False killer whales try to stay together during mass strandings \citep{por77} and drive fisheries (TK pers. comm.). Thus both strandings and drive fisheries tend to include entire social groups, as has been assumed here.

Patterns of growth were found to be similar between the two samples, but both sexes were 10$\%$--20$\%$ larger in Japan than South Africa \citep{fer14} . Additionally, initial ovulation and apparent pregnancy rates were lower in the animals from South Africa, possibly because of impaired reproductive performance in the stranded school. However, it should be noted that there are no juveniles in the samples from: (1) the Japanese shore-drive fisheries (which do not seem to be reproductively compromised) and the (2) stranded sample studied by \cite{alo99} from Tierra del Fuego. Thus, it is likely that there is a mechanism behind the lack of these animals other than diminished reproductive capacity. Survival rates were not calculated separately but the overall age compositions did not suggest any great differences in longevity or survival (see Fig. 2 in \citealp{fer14}). In order to increase sample size, data from South African and Japanese females have been combined for most of the following sections. 

\subsection*{Field procedures}
The procedures used in collecting material in both localities were described by Ferreira \textit{et al.} \citep{fer14}. Attempts were made to collect 1 to 3 adjacent teeth from the center of the lower jaw of each animal. Teeth were fixed in 10\% buffered formalin (Japan) or in 70\% ethanol (South Africa). Where possible, the depth of the mammary gland was measured and the presence/absence of milk recorded, the diameters of both uterine cornua measured and the length and sex of any foetus present recorded. Samples of mammary gland and uterus were taken for histology where appropriate. Both ovaries were collected and the presence of corpora lutea, corpora albicantia or large follicles recorded before the ovaries were fixed in 10\% buffered formalin.

\subsection*{Laboratory procedures}
\subsubsection*{Age determination from teeth}
Counting growth layer groups (GLGs) in tooth sections is the accepted methodology for cetacean age estimation as explained by \cite{per80} and \cite{hoh09}. False killer whale teeth were sectioned longitudinally through the center of the pulp cavity to a thickness of 40-50$\mu$m \citep{fer14}. Sections were then decalcified and stained with haematoxylin before mounting in Canada Balsam. Whales were aged by counting the growth layers in dentine and/or cementum at a magnification of 20-100x (and without reference to other biological data). Growth layer groups in the dentine and cementum were assumed to be deposited annually \citep{kas84b}. The median values of 3 independent GLG counts in the dentine and cementum were taken, and where discrepancies between dentinal and cemental counts occurred, the growth layers in both tissues were repeatedly checked until a good agreement was reached between the two counts. The ages of older individuals with closed pulp cavities were determined using cemental GLG counts only. The ages of individuals below 10 years were estimated to the nearest 0.25 year by comparing the thickness of the first and last postnatal dentinal layers, while in older whales the ages were determined to the nearest $n$ $\pm$ 0.5 year (where $n$ is integer). The estimates for animals aged 10, 20, 40, and 60 were 95\% accurate to within $\pm$0.9, $\pm$1.8, $\pm$2.6, and $\pm$3.4 years, respectively \citep{kas84b}. All further data analyses relate to these age estimates, though data have been grouped where necessary. The life table was constructed with 1 year intervals.

\subsubsection*{Analysis of reproductive tissue: ovaries and mammary glands}
The medulla and cortex of all ovaries were hand-sliced at 1-2 mm intervals and examined macroscopically for various indices of follicular development (non-atretic follicles $<$ 1 mm in diameter), ovulation (corpora lutea of ovulation - CLO, corpora albcantia), pregnancy (corpora lutea of pregnancy - CLP), and follicular atresia (atretic Graafian follicles, corpora atretica). The numbers of corpora lutea, corpora albicantia, and corpora atretica were counted. The diameters of all corpora and Graafian follicles were measured to the nearest 0.1 mm on 3 planes using vernier calipers, and the mean taken as the cube root of the product of the 3 values. Corpora albicantia were classified as young, medium or old according to the characteristics used in the literature \citep{mar84}. Macroscopically visible Graafian follicles (i.e. those $>$ 1 mm in diameter) were classified as atretic or non-atretic on the basis of the macroscopic thickness of the follicle walls. Histological examination of the ovaries was undertaken to confirm macroscopic observations and reproductive status. A detailed description of the terminology is provided by \cite{per84} (see their Appendix A). 

As explained by \cite{per84}, there is considerable evidence that corpora albicantia persist throughout life in at least some cetaceans, as a record of ovulation events. The most comprehensive evidence of persistence comes from the short-finned pilot whale. \cite{mar84} studied the rate of regression in short-finned pilot whales and concluded that some corpora albicantia regressed fully within two years but that rate of regression probably varied with hormonal status. \cite{fer14} also concluded that old corpora albicantia likely persist in the ovaries of false killer whales throughout life (except perhaps for some resorption in the oldest females) and that young and medium corpora albicantia represent progressive stages of regression, the rate of which has not been studied in false killer whales.  

\subsection*{Classification of individuals into reproductive categories}
Following \cite{kas84a} and \cite{mar84}, false killer whale females were classed as mature if they: (a) contained at least 1 corpus luteum or corpus albicans in the ovaries, and$\backslash$or (b) were pregnant or lactating. Mature females were further classified into 5 reproductive categories: pregnant, lactating, ovulating, resting and postreproductive. Pregnancy was determined either by (1) the visible presence of a foetus (or knowledge of its abortion), or (2) the presence of a corpus luteum and evidence from endometrial histology, or (3) the presence of a fragment of placenta or umbilical cord, suggesting that the female had been carrying a foetus or had recently aborted at the time of death. Lactating females were those with: (1) field observations of milk in the mammary gland, or (2) in which histology suggested that the gland was active \citep{fer14}. Ovulating females were those with an active corpus luteum in the ovaries but no signs of pregnancy. The size distribution of these corpora lutea of ovulation tended to be bimodal, with some having diameters $\geq$ 29.6 mm and others diameters $\geq$ 39.3 mm. The latter group coincided with the size range for known corpora lutea of pregnancy and it is possible that some of the larger corpora lutea of ovulation may have been undiagnosed corpora lutea of pregnancy (perhaps associated with very small undetected embryos). Females with at least 1 corpus luteum or corpus albicans in the ovaries but no signs of pregnancy or lactation were classified as resting. Following \cite{mar84}, mature females with no corpora lutea, young or medium corpora  albicantia, or macroscopic follicles in their ovaries were classified as postreproductive.                                                                                                                                                                                                                                                                                                                                                                                                                                                                                                                                                                                                                                                                                                                                                                                                                                                                                                                                                                                                       

\subsection*{Statistical Analysis}
\subsubsection*{Trends in reproductive materials}
Weights of both ovaries were available for 55 females from Japan. Ovary weights of non-pregnant, non-ovulating females (i.e. without an active corpus luteum) were expressed as a proportion of estimated body weight (g/kg). Ovarian weights were averaged and grouped into 8 age classes due to the sparseness of the data (Fig. \ref{fig:fig1}). Age-classes were 5-years wide except for the first. The youngest individual in the dataset was 8.25 years old, so the first category included animals between the ages of 8.25 and 14.5. Similarly, the last category included animals aged 51.0 years or older, with the oldest animal aged at 63.5 years, since there were only two individuals older than 59.5 years. The resulting age-classes were as follows; 8.25-14.5 year-olds, 15.0-20.5 year-olds, 21.0-26.5 year-olds, 27.0-32.5 year-olds, 33.0-38.5 year-olds, 39.0-44.5 year-olds, 45.0-50.5 year-olds and individuals aged 51.0 or older. This age grouping configuration was used in all analyses of reproductive tissues. All statistical analysis was carried out in \texttt{R} (R Core \citealp{rco16}).

We used linear regression models to examine (1) trends in ovarian weights per kg of body mass as a function of age class; (2) trends in mammary gland thickness in different reproductive categories (lactating vs non-lactating) and with age in years; (3) trends in the number of Graafian follicles and the percentage of those that were atretic as a function of age class; and (4) to test for a trend between the total number of corpora albicantia and age in years. In all cases, age was determined by GLGs, as described. We were able to (5) compare the number of corpora lutea representing pregnancy and ovulation in only 2 age classes: animals younger than 25 years and older than 25 years, using a test for proportions. This was because of the very small sample size of 26 individuals. We ran power analyses on all models using the \texttt{pwr} package in \texttt{R} \citep{pwr15}.

\subsubsection*{Construction of the life table}
The calculation of postreproductive representation (PrR) requires data on age-specific fecundity and survival, as well as the cohort size in each age class. The parameters of the life table were estimated from the data for 1 year age classes \citep{cau77, ska10}: 
\vspace{1em}

\begin{adjustwidth}{2.5em}{0pt} 
$fx$: the observed frequency of animals of age $x$ \\
$f'x$: the smoothed frequency of animals of age $x$ \\
$Fx$: the observed cumulative frequency of animals that survived until their $x$ birthday \\
$F'x$: the smooth cumulative frequency of animals that survived until their $x$ birthday \\
$lx$: the survivorship or the probability of surviving to the exact age $x$ \\
$dx$: the frequency of mortality or the probability of dying during the age interval $x$, $x$+1 \\
$qx$: the mortality rate or the proportion of animals alive at age $x$ that die before age $x$+1 \\ 
$px$: the survival rate or the proportion of animals at age $x$ that survive to age $x$+1 
\end{adjustwidth}

\vspace{1em}

Age specific survival and fecundity were derived for animals ranging from age 8 to 63 years old. Younger animals do not appear in the life table because they were not sampled. We calculated age specific life expectancy from the fitted survival data by taking the inverse of a matrix with survival on the diagonal and zeros everywhere else \citep{cas01}. 

Age data were available from a combined data set of 91 females from South Africa and Japan. Data on the age structure of each of the populations separately are not available as a result of the death of PBB in April 2015. To avoid violating the requirements of a vertical life table (i.e. that the frequency of each age class $x$ is equal to or greater than $x$+1; \citep{cau77}), the frequency distribution of animals in the different age classes was smoothed using a regression model before constructing the life table. We used a Generalized Linear Model (GLM) with a Poisson error distribution and a log link function, with counts of animals in each age class as the response variable, and age as an explanatory variable. To find the best model for the age frequency distribution, it is recommended to sequentially add polynomial terms of age \citep{cau77} (i.e., age plus age squared plus age cubed etc) and assess the adequacy of the model via the F-statistic. However, in the case of our data, the only model that yielded a monotonically declining age distribution was the model with age as a linear term. This approach provided values for all ages. 

While survival information was available from a dataset with 1 year age classes, fecundity information had to be pooled into wider age classes. Age classes for fecundity were the same as those used to examine the trends in reproductive materials; 8.25-14.5 year-olds, 15.0-20.5 year-olds, 21.0-26.5 year-olds, 27.0-32.5 year-olds, 33.0-38.5 year-olds, 39.0-44.5 year-olds, 45.0-50.5 year-olds and individuals aged 51.0 or older. This approach boosted the number of data points within each category and made them roughly equal. The effects of different grouping regimes were investigated during exploratory analysis and found to be small. Because the fecundity data were binned into age classes 5 years wide, we fitted a model with age as the explanatory variable in order to obtain values at each age class in the life table. Smoothing of a variable (here fecundity) that is in reality continuous, but sampled over an interval is common practice.

It was not clear what parametric form would be appropriate to model fecundity so we applied a non-parametric, data-driven regression model to the fecundity data to estimate the pregnancy rate for each age class in the survival data. This model was implemented with the \texttt{sm} package in \texttt{R} \citep{bow14}. It was also unclear what the appropriate degrees of freedom should be in this model, so we fitted the model for 10 different values for the degrees of freedom, including the lowest possible and highest possible number ($df$=2.1, 2.9, 3.7, 4.5, 5.3, 6.1, 6.9, 7.7, 8.5, 8.9), and used them in the following steps of the analysis. At the lower end of the range the model fits a parabola, and at the upper end of the range it interpolates between data points, assuming nothing about the shape of the underlying curve. In the absence of information about what this curve ought to look like, we favoured the model with the highest degrees of freedom because it was closest to the data.  

\subsection*{Calculating postreproductive representation (PrR)} 
\label{sec:prr} 
PrR \citep{lev11} is equal to the ratio $T_M/T_B$, where $T_M$ is the expected number of postreproductive years lived by an average newborn, and $T_B$ the expected number of adult years lived by an average newborn. $T_M$ is defined as the product of the remaining life expectancy once 95\% of lifetime fecundity has been realised (age $M$) and the number of individuals surviving to that exact age. $T_B$ is defined as the product of the remaining life expectancy once 5\% of lifetime fecundity has been realised (age $B$) and the number of individuals surviving to that exact age.

Even though the inputs and calculations for PrR are described in terms of expected demographic rates for newborn animals, this measure is independent of infant and juvenile mortality but dependent on survival through the reproductive and postreproductive phases \citep{lev11}, making the method particularly suitable for comparing populations with different pre-reproductive mortality. Based on this premise, we assume that the theory and application are equally relevant in cases where the demographic information is missing for the first few cohorts, as in the dataset analysed here.

One of the advantages of PrR is that it makes it possible to make inferences about the value of PrR obtained for a given species or population. Together, the proportion of animals that were alive in each age class and smoothed fecundity information for false killer whales were used to as inputs to estimate PrR for our sample. We used code provided by Levitis and Lackey \citep{lev11} to obtain a null distribution for the values of PrR that might be expected to obtain based on these data, if the null hypothesis of no postreproductive lifespan were true. We ran the simulation for 1000 populations with the same number of individuals as in the observed dataset ($n$ = 91) and compared the observed PrR to the mean of the null distribution of PrR. The test statistic for the difference between the 2 estimates was obtained by taking the difference of the means and dividing it by the standard deviation of the simulated values. 

Under this methodology, there is no natural way to produce an estimate of the uncertainty associated with the point estimate of PrR for a life table, because it comes from a single population. The life table for false killer whales was constructed based on combined data from two populations (a healthy population from the six harvested schools from Japan, and a population with possibly impaired vital rates from one stranded school from South Africa). We could only obtain survival information for the combined dataset, as explained above, but were able to calculate fecundity separately. This approach made it possible to calculate point estimates and null distributions for PrR based on the combined dataset and for the data from South Africa and Japan separately. In the absence of a formal measure of uncertainty in PrR, we present the 2 separate point estimates as an indication of the range that PrR can take on for false killer whales under widely differing scenarios for reproductive rates. In addition, we illustrate the effect of the smoothness of the fecundity curve (with 10 different values, starting from the most smooth and ending on the least smooth) and the resulting PrR estimate.

\section*{Results}

\subsection*{Ovarian weights}
Based on the weights of both ovaries from 55 females, we found no evidence of age-related changes in the ovarian size (ovary weight per kg of body weight) of false killer whales. However, there was very low power to detect a significant effect ($F_{7, 47}$ = 0.88, $p$-value for the difference in ovary weight in any 2 groups being substantially different from zero equal to 0.53). It appears that ovary weights increased with age up to about 26 years as expected (Additional file 1) (see \citealp{mar84} for short-finned pilot whales), but this trend was not statistically significant. With our sample size there was a 11\% chance of detecting an effect of the observed size (0.03) at the 5\% level if it was actually there.

\subsection*{Mammary gland development with age}
Measurements of mammary gland thickness were available for 38 mature females and averaged 2.5 cm (mean, range 0.9-4.2 cm). There was strong evidence that the gland in lactating females (mean thickness 3.12 cm) was significantly thicker than that in mature, non-lactating females (mean thickness 2.13 cm) (\cite{fer14} and this study: $F_{2, 29}$ = 7.04, $p$-value for no difference in mammary gland thickness between lactating and non-lactating individuals $<$ 0.01). There was sufficient power to conclude that this result was robust for the observed effect size and sample size (99\% chance of detecting the effect at the 5\% level). However, there was no evidence to support a decline in mammary gland thickness with age (Additional file 2), despite reasonable power to detect a significant effect of the observed size with the sample size (78\% chance of detecting an effect of size 1 at the 5\% level).

\subsection*{Follicular development and atresia with age}
\begin{figure}[!ht]
  \centering \includegraphics[width=0.7\textwidth]{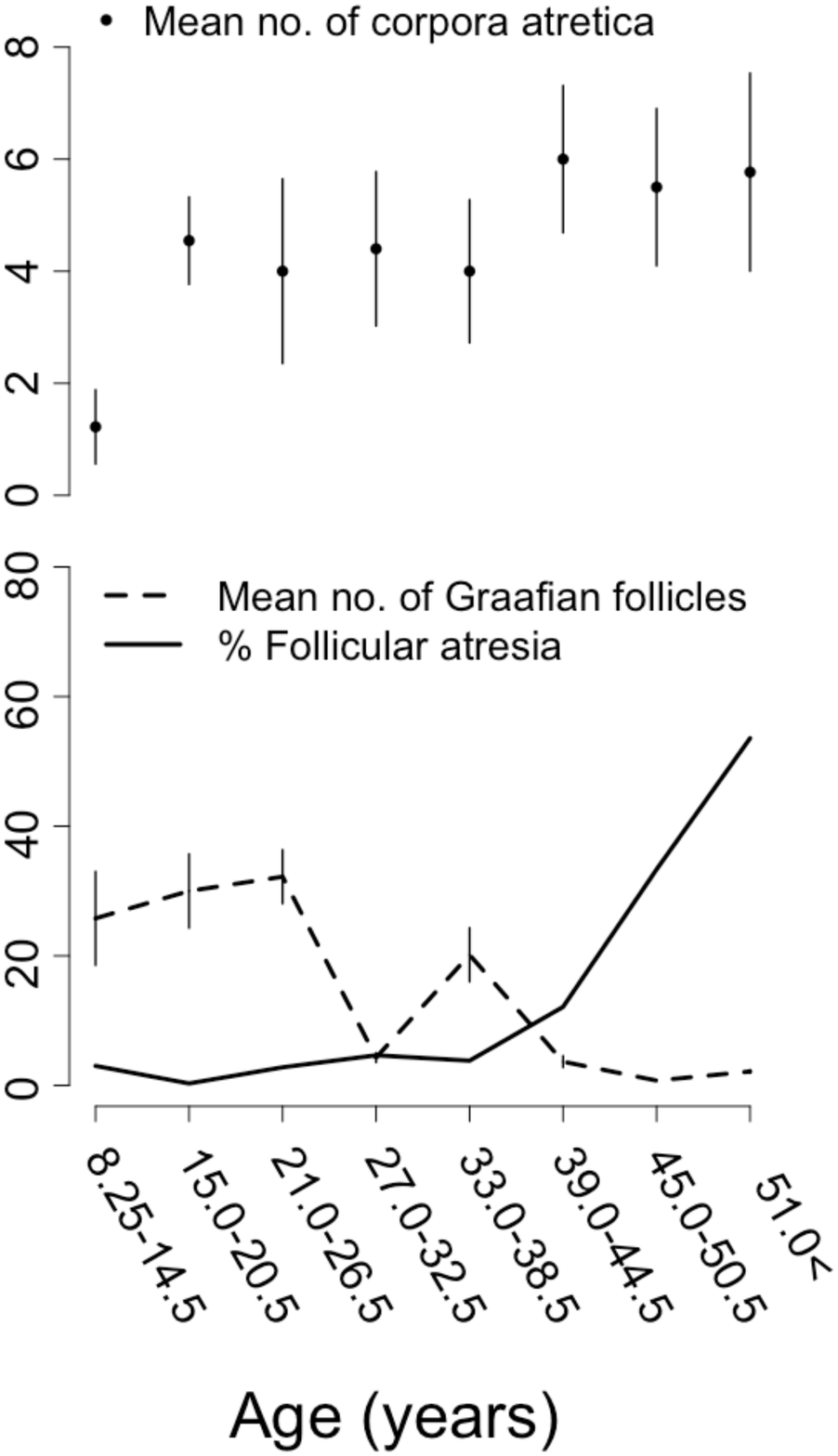}
  \caption{The trend with age in the mean number of corpora atretica (upper plot: mean number per whale in age class $\pm$ standard error of the mean), the number of Graafian follicles (lower plot broken line: mean number per whale in age class $\pm$ standard error of the mean), the percentage of Graafian follicles that were atretic (lower plot solid line: mean percentage per whale in age class) in false killer whales from South Africa and Japan ($n$=38).}
      \label{fig:fig1}
      \end{figure}

Follicle counts were available for 83 individuals. The number of macroscopically-visible Graafian follicles in mature females varied greatly between individuals, with 42.2$\%$ (35 out of 83) having none. The number of Graafian follicles remained on average high in the first 3 age groups (8.25-26.5 year-olds: median 1, mean 29.5$\pm$10.12 SE), but above 27 years of age the number declined markedly (17.0-51.0+ year-olds: median 1, mean 7.01$\pm$3.77 SE) (Fig. \ref{fig:fig1} dashed line). Despite the spike in Graafian follicles in the age group including 33.0-38.5 year-olds, there was evidence for an overall decrease in Graafian follicles with age ($F_{1, 81}$ = 5.446, $p <$ 0.03) and there was sufficient power to conclude that this trend was robust (99$>$\% chance of detecting an effect of size 4.4 at the 5\% level). 

Follicular atresia gradually increased with age, but not markedly in older individuals. The number of corpora atretica per female showed a progressive increase with age ($F_{1, 81}$ =  5.06, $p <$ 0.03), as did the mean percentage of follicles that were atretic, resulting from the atresia of luteinized Graafian follicles (Fig. \ref{fig:fig1} solid line). There was sufficient power to judge this result to be robust (99\% chance of detecting an effect of 0.5 at the 5\% level). 

\subsection*{Regression of corpora albicantia with age}
\begin{figure}[!ht]
  \centering \includegraphics[width=0.7\textwidth]{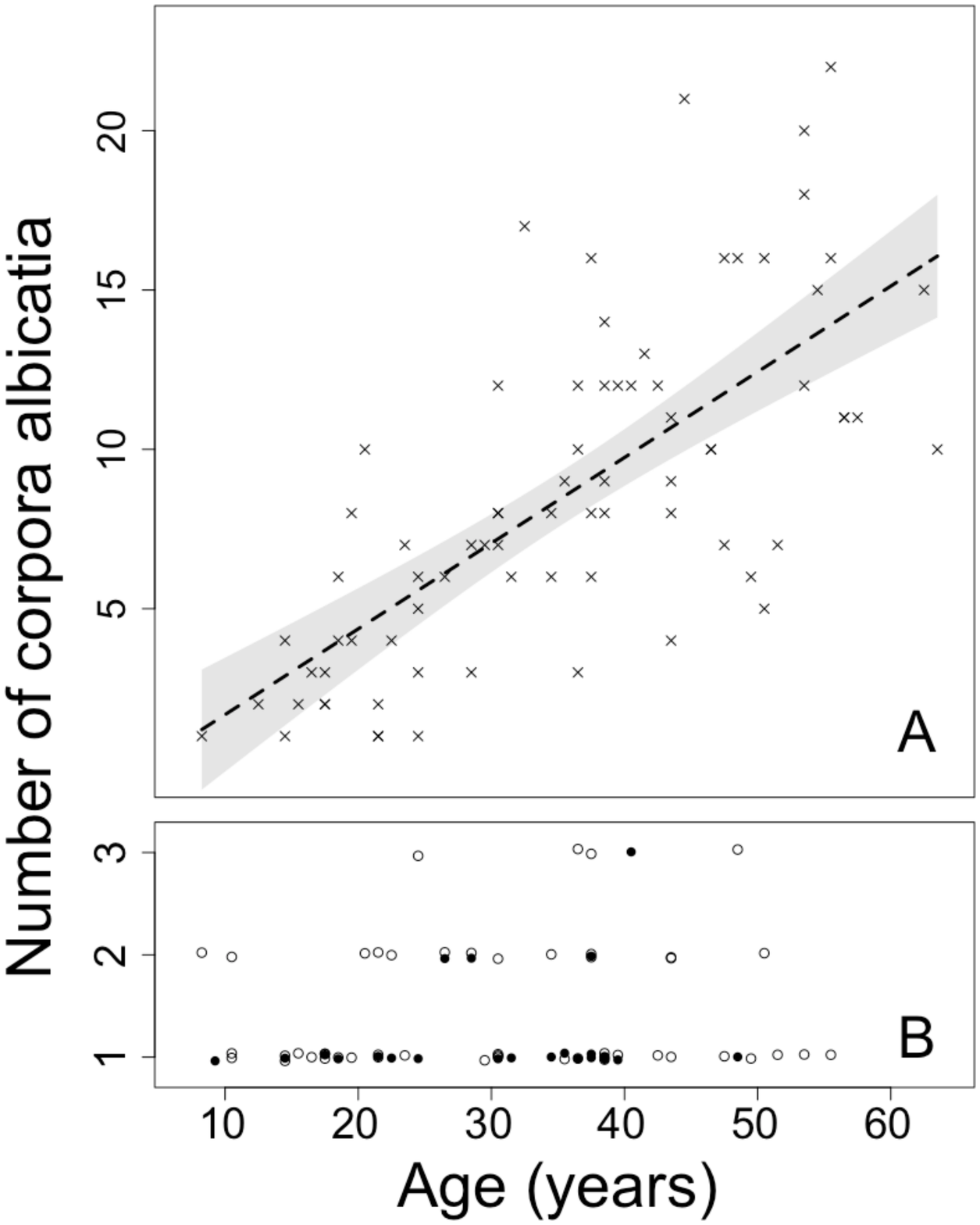}
  \caption{Numbers of old corpora albicantia (A), and medium (open circles) and young (solid circles) corpora albicantia (B) in 102 mature female false killer whales of different ages from Japan and South Africa. The count of corpora albicantia has been jittered in (B) for plotting purposes.}
      \label{fig:fig2}
      \end{figure}

Data on corpora albicantia were available from 75 individuals. There was strong evidence that the numbers of old corpora albicantia in an individual's ovaries continued to increase linearly with age ($F_{1, 73}$ = 75.4, $p <$ 0.01) (Fig. \ref{fig:fig2} A) and there was enough power to conclude this trend was robust ($>$99\% chance of detecting an effect of size 0.3 at the 5\% level). The maximum number of old corpora albicantia was 22 (in an individual 55.5 year old), whereas the maximum number of medium or young corpora albicantia in any individual was only 3 and showed no signs of accumulating except in newly mature (8-9 year old females) individuals (Fig. \ref{fig:fig2} B). Only 2 of the 31 females over 40 years of age had any young corpora albicantia: the oldest female presenting a young corpus albicans was a 48.5 years old resting female and the second oldest female was aged 40.5 years and was ovulating/lactating. None of the 15 females older than 48.5 years contained young corpora albicantia and the oldest female with a medium corpus albicans was 55.5 years old. This pattern suggests the onset of a reduction in the rate of successful ovulation in animals at some point after 41 years of age, although the shape of this trend is not known.

\subsection*{Trend in pregnancy rate and successful ovulations with age}
The relative frequencies of corpora lutea of ovulation and pregnancy were available for 26 known-age female false killer whales using combined data from Japan and South Africa (accessory corpora lutea have been excluded). The data were not sufficient to compare the relative frequencies across the 8 age groups used in the other analyses of reproductive tissue so the data were split into 2 groups: animals up to 25 years old and animals older than 25 years. This cutoff was chosen so that the sample sizes were equal. The frequency of corpora lutea of pregnancy relative to those of ovulation was greater in females up to 25 years of age compared to older females (mean difference 0.38, SE 0.18) (Additional file 3). The relative frequencies were compared in the 2 age groups to test whether ovulation was less likely to be followed by pregnancy in older animals. There was insufficient evidence to support a true difference between the groups ($\chi^2$ = 2.52, $df$=1, $p$ = 0.11) but there was also low power to detect a trend (10\% chance of detecting an effect of size 0.38 at the 5\% level). We would need a sample size of approximately 108 individuals to have an 80\% chance of detecting that same trend at the 5\% level.

\subsection*{Statistical analysis of PRLS}
\subsubsection*{The life table}
\begin{figure}[!ht]
  \centering \includegraphics[width=0.9\textwidth]{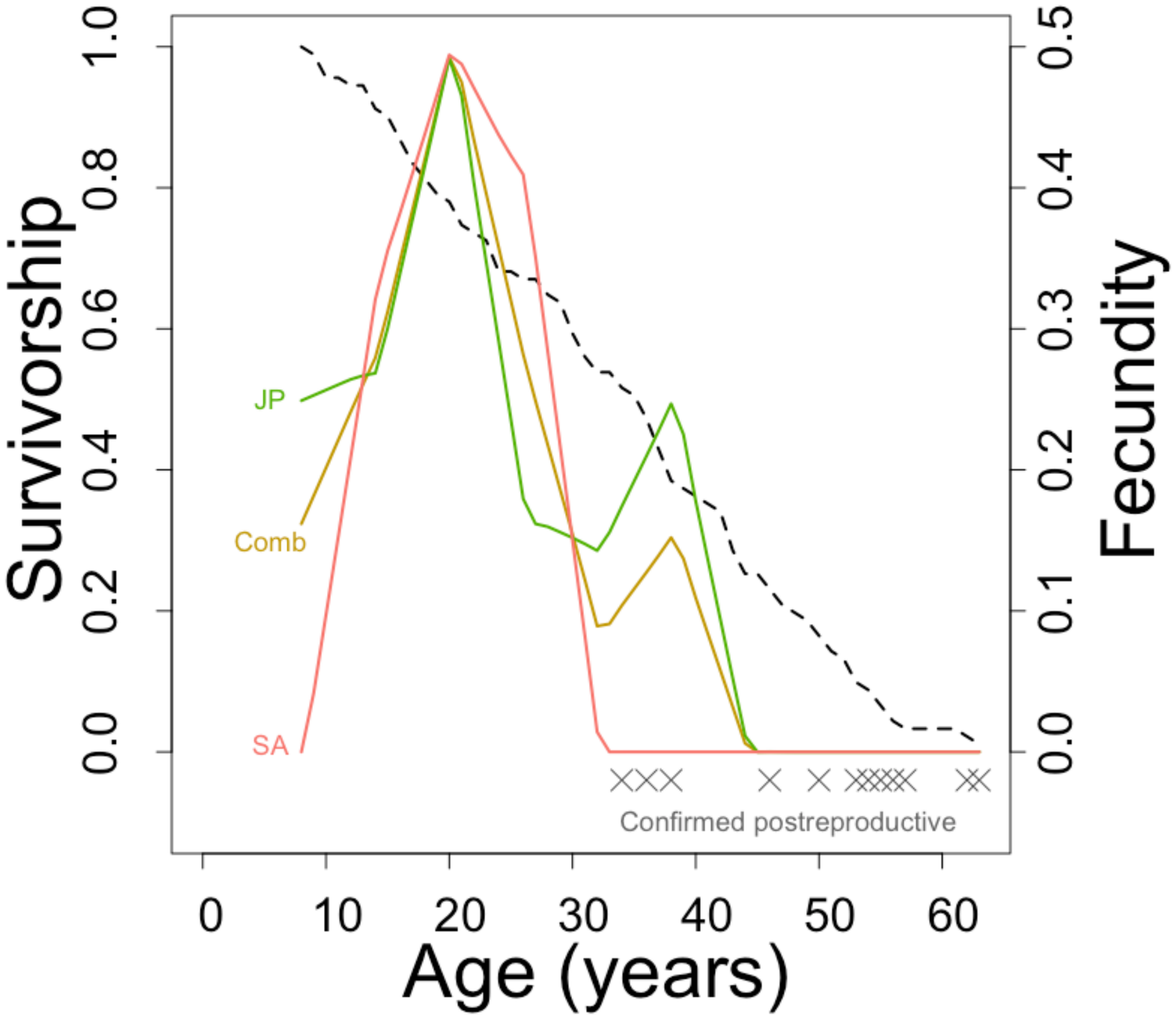}
  \caption{Survivorship and fecundity plots for female false killer whales from the South African, Japanese and combined datasets. Survivorship (proportion of animals surviving to an age) was only estimated for the combined population dataset (dashed black line). The crosses represent females of known age from the combined dataset that were found to be postreproductive according to the criteria defined in the text.}
      \label{fig:fig3}
      \end{figure}

Our vertical life table for false killer whales is provided as Additional file 4. In order to increase sample size, age-specific pregnancy rates were investigated for both the Japanese and South African samples combined (Table \ref{tab:tab1}). Even so, data had to be binned by age group to obtain sufficient resolution (Table \ref{tab:tab1}). The apparent pregnancy rate rose from 16.7$\%$ in newly mature whales to 50$\%$ in females aged 20 - 26 years, declining thereafter to 15.4$\%$ in 38 - 44 years old females. The oldest pregnant female was 43.5 years old. None of the 23 whales 44-63 years old was pregnant. Mean age specific fecundity was estimated at 0.16 (SE 0.06, max. 0.50).  We calculated age-specific fecundity, the total number of offspring of both sexes born to all individuals within an age class \citep{lev11}, as the smoothed proportion of pregnant females using the raw data (Table \ref{tab:tab1}) (Fig. \ref{fig:fig3}). We used the model with the highest degree of smoothing in this case, to follow the data as closely as possible (Additional file 5). 

The fitted values for the separate and combined population data can be found in Additional files 6-8.

\begin{table}[!ht]
\caption{Variation in apparent pregnancy rates of Japanese and South African false killer whales with age. Labels \textit{Pregnant} and \textit{Pregnant$^{a}$} are used to denote the presence of a foetus, and no foetus but corpus luteum $\geq$ 3.93mm thick, respectively (see Materials and Methods).}
      \begin{tabular}{c | ccc | ccc | ccc | ccc}
        \hline
	\multirow{2}{*}{Age class} & \multicolumn{3}{c |}{Pregnant} & \multicolumn{3}{c |}{Pregnant$^{a}$} &\multicolumn{3}{c |}{Total mature} & \multicolumn{3}{c }{Pregnant$\slash$Total} \\ 
	& JP & SA & Both & JP & SA & Both & JP & SA & Both & JP & SA & Both \\ \hline
      8.25 & 1 & 0 & 1 & 0 & 0 & 0 & 4 & 2 & 6 & 0.25 & 0.00 & 0.17 \\ 
      14.25 & 3 & 1 & 4 & 0 & 0 & 0 & 11 & 3 & 14 & 0.27 & 0.33 & 0.29 \\
      20.25 & 3 & 0 & 3 & 1 & 1 & 2 & 8 & 2 & 10 & 0.50 & 0.50 & 0.50 \\ 
	26.25 & 1 & 0 & 1 & 0 & 2 & 2 & 6 & 5 & 11 & 0.17 & 0.40 & 0.27 \\ 
	32.25 & 0 & 0 & 0 & 1 & 0 & 1 & 7 & 5 & 12 & 0.14 & 0.00 & 0.08 \\ 
	38.25 & 2 & 0 & 2 & 0 & 0 & 0 & 8 & 5 & 13 & 0.25 & 0.00 & 0.15 \\ 
	44.25 & 0 & 0 & 0 & 0 & 0 & 0 & 3 & 5 & 8 & 0.00 & 0.00 & 0.00 \\ 
	50.25 & 0 & 0 & 0 & 0 & 0 & 0 & 7 & 3 & 10 & 0.00 & 0.00 & 0.00 \\ 
	56.25 & 0 & 0 & 0 & 0 & 0 & 0 & 3 & 3 & 5 & 0.00 & 0.00 & 0.00 \\ 
	Total & 10 & 1 & 0 & 2 & 3 & 5 & 57 & 32 & 89 & - & - & - \\ \hline
      	\end{tabular}
       \label{tab:tab1}
\end{table}

\subsubsection*{Postreproductive representation (PrR)}
For the combined population ages at which an animal has reached 5$\%$ and 95$\%$ of its reproductive potential were found to be 10 and 39 years respectively (5th and 95th fecundity quantiles), and the age at which an animal has lived 95$\%$ of its lifespan was found to be 51 years. The population-specific values can be seen in Table \ref{tab:tab2}. The exact value of PrR for the combined population data was found to be 0.14. This value was substantially different to the mean of the null distribution for PrR (99\% CI 0.0003--0.0826) under the null hypothesis of no post reproductive lifespan Table \ref{tab:tab2}, and no simulated value exceeded 0.13. These results are strong evidence for a postreproductive lifespan in false killer whales. In calculating separate point estimates for each population, representing different fecundity scenarios, we found that the evidence in favour of postreproductive lifespan strengthened when the data from South Africa were considered on their own (PrR = 0.37), and remained effectively unchanged if data from Japan were considered on their own (PrR = 0.12). The test statistic for the difference in the combined dataset was found to be 7.4 standard deviations from zero, and the $p$-value was effectively zero. 

\begin{table}[!ht]
\caption{PrR calculations under different scenarios for the smoothness of the regression model for fecundity, presented for each population separately and for the combined data set. See the text and \cite{lev11} for definitions.}
\resizebox{\textwidth}{!}{%
\begin{tabular}{c | cc | cc | cc | cc | cc | cc}
  \hline
\multicolumn{13}{c}{\multirow{2}{*}{SEPARATE POPULATION ANALYSES}} \\
\multicolumn{13}{c}{\multirow{2}{*}{  }} \\
\hline
Smoothing function & \multicolumn{2}{c |}{Age B} & \multicolumn{2}{c |}{Age M} & \multicolumn{2}{c |}{TB} & \multicolumn{2}{c |}{TM} & \multicolumn{2}{c |}{Japan} & \multicolumn{2}{c }{South Africa}\\ 
degrees of freedom & JP & SA & JP & SA & JP & SA & JP & SA & PrR & $H_0$ 99\%CI & PrR & $H_0$ 99\%CI \\ \hline
2.1 & 9 & 9 & 45 & 38 & 18.96 & 18.96 & 1.09 & 2.78 & 0.06 & 0.00--0.11 & 0.15 & 0.00--0.11 \\ 
2.9 & 9 & 9 & 43 & 37 & 18.96 & 18.96 & 1.49 & 3.09 & 0.08 & 0.00--0.11 & 0.16 & 0.00--0.11 \\ 
3.7 & 9 & 10 & 43 & 35 & 18.96 & 18.14 & 1.49 & 3.76 & 0.08 & 0.00--0.09 & 0.21 & 0.00--0.11 \\ 
4.5 & 9 & 11 & 43 & 33 & 18.96 & 17.33 & 1.49 & 4.49 & 0.08 & 0.00--0.10 & 0.26 & 0.00--0.11 \\ 
5.3 & 9 & 12 & 42 & 32 & 18.96 & 16.55 & 1.72 & 4.88 & 0.09 & 0.00--0.09 & 0.29 & 0.00--0.11 \\ 
6.1 & 9 & 12 & 42 & 31 & 18.96 & 16.55 & 1.72 & 5.29 & 0.09 & 0.00--0.06 & 0.32 & 0.00--0.10 \\ 
6.9 & 9 & 12 & 41 & 30 & 18.96 & 16.55 & 1.96 & 5.72 & 0.10 & 0.00--0.04 & 0.35 & 0.00--0.09 \\ 
7.7 & 9 & 12 & 40 & 29 & 18.96 & 16.55 & 2.22 & 6.17 & 0.12 & 0.00--0.05 & 0.37 & 0.00--0.09 \\ 
8.5 & 9 & 12 & 40 & 29 & 18.96 & 16.55 & 2.22 & 6.17 & 0.12 & 0.00--0.04 & 0.37 & 0.00--0.04 \\ 
8.9 & 9 & 12 & 40 & 29 & 18.96 & 16.55 & 2.22 & 6.17 & 0.12 & 0.00--0.03 & 0.37 & 0.00--0.03 \\ 
\hline \hline
\multicolumn{13}{c}{\multirow{2}{*}{COMBINED POPULATION ANALYSIS}} \\
\multicolumn{13}{c}{\multirow{2}{*}{  }} \\
\hline
Smoothing function & \multicolumn{2}{c |}{\multirow{2}{*}{Age B}} & \multicolumn{2}{c |}{\multirow{2}{*}{Age M}} & \multicolumn{2}{c |}{\multirow{2}{*}{TB}} & \multicolumn{2}{c |}{\multirow{2}{*}{TM}} & \multicolumn{2}{c |}{\multirow{2}{*}{PrR}} & \multicolumn{2}{c }{\multirow{2}{*}{$H_0$ 99\% CI}} \\ 
degrees of freedom & & & & & & \\
  \hline
2.1 & \multicolumn{2}{c |}{9} & \multicolumn{2}{c |}{42} & \multicolumn{2}{c |}{18.96} &\multicolumn{2}{c |}{ 1.72} & \multicolumn{2}{c |}{0.09} & \multicolumn{2}{c }{0.00--0.11} \\ 
2.9 & \multicolumn{2}{c |}{9} & \multicolumn{2}{c |}{41} & \multicolumn{2}{c |}{18.96} & \multicolumn{2}{c |}{1.96} & \multicolumn{2}{c |}{0.10} & \multicolumn{2}{c }{0.00--0.11} \\ 
3.7 & \multicolumn{2}{c |}{9} & \multicolumn{2}{c |}{41} & \multicolumn{2}{c |}{18.96} & \multicolumn{2}{c |}{1.96} & \multicolumn{2}{c |}{0.10} & \multicolumn{2}{c }{0.00--0.11} \\ 
4.5 & \multicolumn{2}{c |}{10} & \multicolumn{2}{c |}{40} & \multicolumn{2}{c |}{18.14} & \multicolumn{2}{c |}{2.22} & \multicolumn{2}{c |}{0.12} & \multicolumn{2}{c }{0.00--0.10} \\ 
5.3 & \multicolumn{2}{c |}{10} & \multicolumn{2}{c |}{40} & \multicolumn{2}{c |}{18.14} & \multicolumn{2}{c |}{2.22} & \multicolumn{2}{c |}{0.12} & \multicolumn{2}{c }{0.00--0.10} \\ 
6.1 & \multicolumn{2}{c |}{10} & \multicolumn{2}{c |}{40} & \multicolumn{2}{c |}{18.14} & \multicolumn{2}{c |}{2.22} & \multicolumn{2}{c |}{0.12} & \multicolumn{2}{c }{0.00--0.10} \\ 
6.9 & \multicolumn{2}{c |}{10} & \multicolumn{2}{c |}{39} & \multicolumn{2}{c |}{18.14} & \multicolumn{2}{c |}{2.49} & \multicolumn{2}{c |}{0.14} & \multicolumn{2}{c }{0.00--0.10} \\ 
7.7 & \multicolumn{2}{c |}{10} & \multicolumn{2}{c |}{39} & \multicolumn{2}{c |}{18.14} & \multicolumn{2}{c |}{2.49} & \multicolumn{2}{c |}{0.14} & \multicolumn{2}{c }{0.00--0.09} \\ 
8.5 & \multicolumn{2}{c |}{10} & \multicolumn{2}{c |}{39} & \multicolumn{2}{c |}{18.14} & \multicolumn{2}{c |}{2.49} & \multicolumn{2}{c |}{0.14} & \multicolumn{2}{c }{0.00--0.07} \\ 
8.9 & \multicolumn{2}{c |}{10} & \multicolumn{2}{c |}{39} & \multicolumn{2}{c |}{18.14} & \multicolumn{2}{c |}{2.49} & \multicolumn{2}{c |}{0.14} & \multicolumn{2}{c }{0.00--0.08} \\ 
  \hline
\end{tabular}}
\label{tab:tab2}
\end{table}

\begin{figure}[!ht]
  \centering \includegraphics[width=0.8\textwidth]{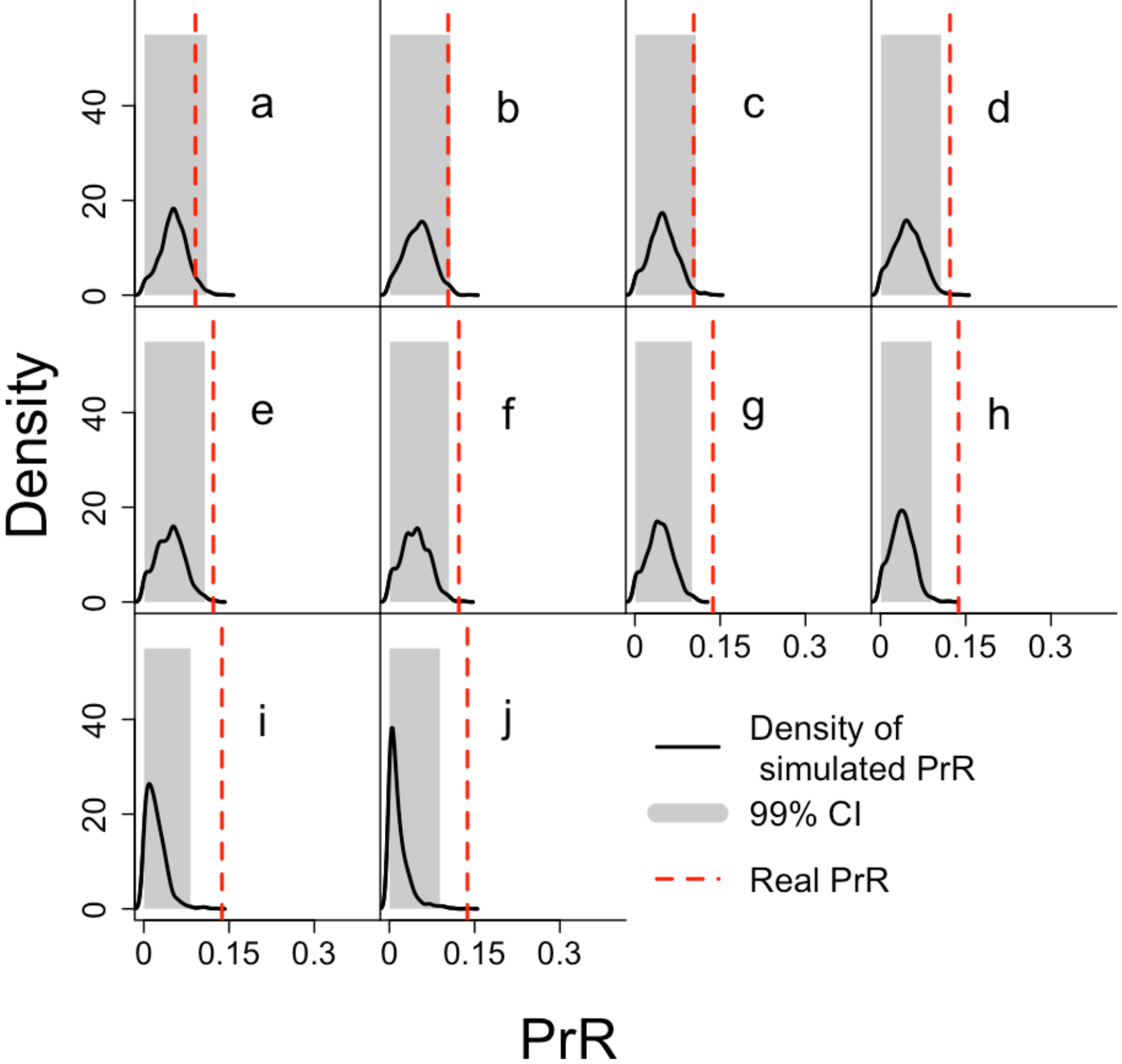}
  \caption{The simulated distribution of PrR under the null hypothesis of no PRLS (black curve) with its 99\% quantile shaded in grey, and the estimated value for PrR, under 10 scenarios for fecundity. Each plot illustrates the results of the PrR analysis with fecundity smoothed using data-driven regression with a different number of degrees of freedom, from lowest to highest.}
      \label{fig:fig4}
      \end{figure}

The null distribution of the simulated PrR's and the estimated PrR for the combined population are shown in Fig. \ref{fig:fig4} for the 10 different smoothing scenarios for fecundity shown in Additional file 5. Equivalent representations for the Japanese and South African populations on their own are shown in Additional file 9. For the combined population data, the estimate of PrR was outside of the 99\% CI of the null distribution for no PRLS for 7 out of the 10 models, when the smoothing was carried out with degrees of freedom equal to 4.5 or greater. When the Japanese data were considered on their own the estimate of PrR was outside the 99\% CI for 5 out of the 10 models, when the degrees of freedom for the smooth were equal to 6.1 or greater. When the South African data were considered on their own, the estimate of PrR was outside the 99\% CI for the null distribution in all 10 models. 

Our sample included 15 postreproductive females based on the ovarian status as defined above. The ages of the 14 animals for which an age estimate was available ranged from 34.5 to 63.5 (mean 50.9 $\pm$ 2.5 SE) years, with 50\% of whales over 45 years, and all whales over 55 years old being postreproductive (Fig. \ref{fig:fig3}). The maximum ages of whales in both samples were similar, 63.5 and 62.5 years, in females from South Africa and Japan respectively.

\section*{Discussion}
The morphological and statistical evidence we have presented makes a strong case for the false killer whale being the third species of toothed whale from the family Delphinidae to have a prolonged postreproductive lifespan (PRLS). 
The morphological evidence includes: (1) the marked decline in macroscopically visible Graafian follicles in the ovaries of females above 38.5 years of age, and complete absence of such follicles from the ovaries of females over 57.5 years of age; and (2) the apparent reduction in the rate of successful ovulation in animals in their fifth and sixth decades of life as evidenced by the lack of young corpora albicantia in their ovaries, a pattern suggesting an absence of ovulation in animals over 49 years of age. There was a suggestion of an age-related decline in the ratio of corpora lutea of pregnancy to corpora lutea of ovulation, which would suggest that ovulation is less likely to be followed by pregnancy in older females, but the sample size was too small to show this convincingly.

The statistical evidence for a prolonged PRLS in false killer whales is also compelling, indicating that the PrR for the combined sample was 0.14, substantially different to the mean of the null distribution under the null hypothesis of no PRLS. Evidence supporting a substantial PRLS strengthened when the data from South Africa were considered on their own (PrR = 0.37), and remained effectively unchanged when the Japanese data were considered on their own (PrR = 0.13). Even though the South African sample may have had impaired fecundity, it did not bias the PrR estimate for the combined estimate.

Nonetheless, the relatively few young individuals (both mature and immature) in our false killer whale sample is a shortcoming, because the calculation of PrR required that all age classes were represented. However, it seems unlikely that young, sexually mature individuals were grossly underrepresented in our samples, based on the youngest mature animals present in the sample (South Africa) and the estimated age at which 50\% of animals are mature (Japan) and the propensity of false killer whale groups to stay together. 

The fact that the data from South Africa are from only 1 group limits robust population comparisons. Genetic sampling indicates considerable population structure \citep{chi07, chi10} in false killer whales. It will thus be important to collect additional data from stranded animals to test if the results reported here apply to other social groups and populations which may be subject to differed selection pressures. False killer whales strandings tend to involve large groups \citep{bai08} and these mass stranding events should be recognised as opportunities to collect further life history data. 

\begin{figure}[!ht]
  \centering \includegraphics[width=1\textwidth]{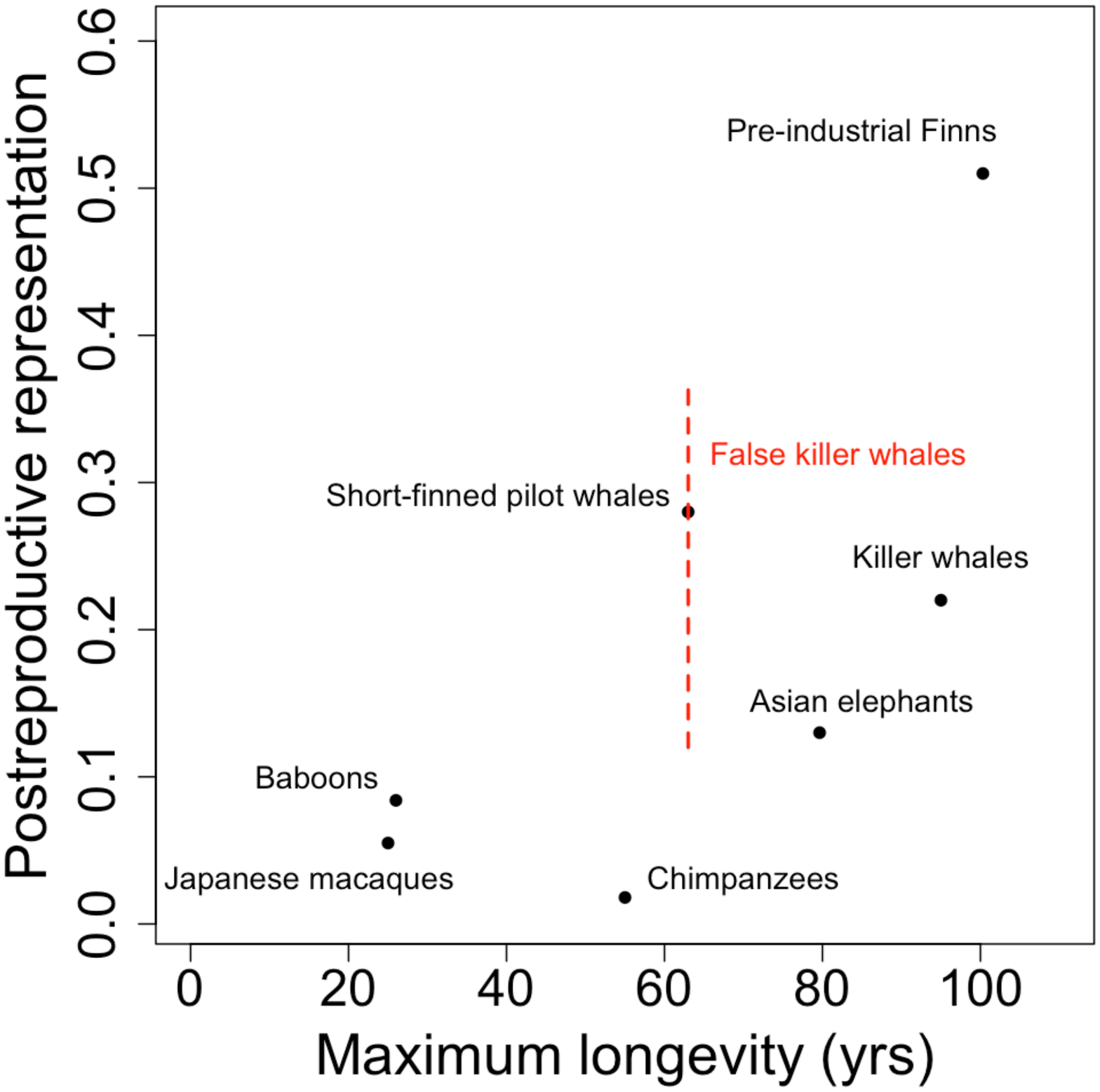}
  \caption{Postreproductive representation (PrR) plotted against maximum longevity in false killer whales and other large mammals. This figure was adapted from Fig. 5 in \cite{lah14} to include the estimate for false killer whales from this study. PrR values and maximum longevities in Asian elephants and humans from \cite{lah14}, and in killer whales from \cite{cro15}. Other PrR-values were taken from \cite{lev11}. Maximum longevities in Japanese macaques were sourced from \cite{pav99}, chimpanzees from \cite{hil01}, baboons from \cite{bro02} and short-finned pilot whales from \cite{kas84a}, as in the original version of the figure.}
      \label{fig:fig5}
      \end{figure}

PrR has been calculated for several species of large mammals \citep{lev11, lah14}. Our results for the false killer whale span the single estimates for the other cetaceans that are accepted as having a significant PRLS (0.22 for killer whales and 0.28 for short-finned pilot whales), are similar to the values for working Asian elephants \textit{Elephas maximus} (0.13) and higher than for 3 long-lived, non-human primates living in wild or semi-wild conditions (chimpanzees \textit{Pan troglodytes} (0.018); Japanese macaques \textit{Macaca fuscata} (0.055); baboons \textit{Papio hamadryas} (0.084)) but lower than pre-industrial humans from Finland (0.51), who PrR estimate corresponds with patterns found in other historical or hunter-gatherer human populations (0.3-0.47) (Fig. \ref{fig:fig5} this study; \citealp{lev11} and \citealp{lah14} for more details). Even if the lower estimate from the Japanese population were to be considered on its own, it is still significantly different from what we would expect under no PRLS.

Both adaptive and non-adaptive hypotheses have been proposed to explain the evolution of PLRS \citep{nic16}. The non-adaptive hypothesis assumes that PLRS primarily results from the non-adaptive ``mismatch'' between somatic and reproductive ageing, a phenomenon that is likely to be more obvious in long-lived species. The adaptive hypotheses assume that PRLS evolved via the fitness benefits gained by helping kin \citep{mca05, war09, fos12, bre15}, especially when local relatedness increases with female age, favouring late life helping \citep{joh10}.  Tests of these hypotheses are rare \citep{lah16, nic16}, largely because of the lack of the required data.

\cite{nic16} used a phylogenetic approach to test the adaptive and non-adaptive hypotheses in some mammalian species. The proportion of life spent in the postreproductive phase was related to lifespan and patterns of philopatry, suggesting that the duration of PRLS may be impacted by both non-adaptive and adaptive processes. The analysis of \cite{nic16} supported the theory that PLRS primarily results from the non-adaptive ``mismatch'' between somatic and reproductive ageing but suggested that patterns of philopatry may subsequently confer adaptive benefits to late-life helping by increasing relatedness within kinship groups with age. The strength of such inferences would be greatly improved by using rigorous morphological and statistical criteria for PLRS such as presented here. 

Longitudinal data are required to test hypotheses about the ways in which a PRLS might increase the inclusive fitness of a false killer whale population, as has been done for 2 killer whales populations off British Columbia and Washington \citep{war09, fos12, bre15}, and working Asian elephants \citep{lah16}. Such data will be logistically challenging to collect for false killer whales \citep{bai08} because of their pelagic lifestyle. The most promising source of such data is the long-term study of the false killer whale population around the Hawaiian Islands \citep{bai08, bai16}. Understanding the way in which local relatedness changes with female age would be particularly informative from a theoretical perspective, especially since it has been demonstrated that an increase in relatedness with female age can arise in species with very different social structures \citep{joh10}. It would also be particularly informative, albeit logistically challenging, to obtain data about changes in female relatedness with age, social structure and evidence for a PRLS in other large, social odontocetes such as sperm whales (\textit{Physeter macrocephalus}), Risso's dolphins (\textit{Grampus griseus}), melon headed whales (\textit{Peponocephala electra}), pygmy killer whales (\textit{Feresa attenuata}) and pilot whales (\textit{Globicephala spp.}). Such data would provide evidence for or against a prolonged PRLS being a generic characteristic of such species, and enable the non-adaptive and adaptive hypotheses for the phenomenon to be further tested. 

\section*{Conclusions}
We found morphological and statistical evidence for PRLS in South African and Japanese pods of false killer whales. Based on our results, the false killer is the third non-human mammal in which this phenomenon has been convincingly demonstrated in wild populations. Our estimates for the PrR of the false killer whale (0.13-0.37; Fig. \ref{fig:fig5}) span the single value available for the short-finned pilot whale (0.28), and killer whale (0.22) and are comparable with estimates for historical or hunter-gather human populations (0.3-0.47). Our results suggest that it would be fruitful to investigate the age-related changes in relatedness and social structure of false killer whales and other large social odontocetes in order to gain further theoretical insights into why some social large mammals exhibit a postreproductive phase in the life history of females.

\section*{List of abbreviations}
CI: confidence interval
CLO: corpus luteum of ovulation
CLP: corpus luteum of pregnancy
GLG: growth layer group
PrR: postreproductive representation
PrT: postreproductive time
PRLS: postreproductive lifespan
SE: standard error

\section*{Ethics approval and consent to participate}
Not applicable.

\section*{Consent for publication}
Not applicable.

\section*{Availability of data and material}

The data set(s) supporting the results of this article are available in the Zenodo repository (permanent link to the data will be added upon article acceptance) and in the article's additional file(s).




\section*{Competing interests}
The authors declare that they have no competing interests.

\section*{Funding}
This research was supported by the Nature Conservation Society of Japan and US Marine Mammal Commission. TP was supported by a Scarce Skills Postdoctoral Research Fellowship from the National Research Foundation, South Africa during part of the time spent working on this manuscript. Financial support for the work in Japan was provided by the World Wide Fund for Nature, Japan, and in South Africa by a grant to PBB from the National Research Foundation, South Africa. 

\section*{Author's contributions}
TP carried out the statistical analysis and wrote much of the manuscript text. IMF carried out the laboratory analyses of the South African sample and contributed to an early draft of the manuscript. PBB collected samples from the South African data, helped conceive the study and contributed to a mature draft of the manuscript. TK contributed samples from the Japanese data, helped conceive of the study and contributed to a mature draft of the manuscript. HM carried out the laboratory analyses of the Japanese sample, helped conceive the study, reviewed the literature and framed much of the biological discussion. All authors read and approved the final manuscript apart from PBB who died in May 2015.

\section*{Acknowledgements}
TK acknowledges the Katsumoto Fishery Cooperative Union for offering the opportunity to study their false killer whale carcasses, and the team of volunteers that assisted withd the collection of samples in 1979 and 1980. IF and PBB thank Graham Ross, Vic Cockcroft and others in the team who assisted with data and sample collection from the 1981 St Helena Bay stranding. HM thanks Savita Francis and Leigh Winsor for assistance with laboratory processing of the ovaries from Japan. IF would also like to acknowledge Rina Owen and Schalk Human, Department of Statistics, University of Pretoria, for statistical advice in the early stages of analysis, and Steven Austad, University of Texas Health Science Center, Robin Baird, Cascadia Research Collective and Stephanie Pl\"{o}n, Port Elizabeth Museum, for valuable comments and suggestions. Annamarie Bezuidenhout and Hannetjie Bruwer, Academic Information Service, University of Pretoria, assisted in procuring references. Robin Baird and two anonymous referees provided valuable comments on earlier versions of the manuscript.

\bibliographystyle{plainnat} 

\section*{Additional Files}
\subsection*{Additional file 1 --- Thickness of mammary gland in false killer whales (South Africa) in relation to age and reproductive status.}

 \subsection*{Additional file 1 --- Mean ovary weights (g per kg of estimated body mass) for non-pregnant, non-ovulating female false killer whales.} \label{sec:adfile1}
Mean weight of both ovaries in grams per kilogram of estimated body mass as a function of age in a sample of non-pregnant, non-ovulating female false killer whales from Japan ($n$ = 55). We found no evidence for a trend in mean ovary weight with age, though the power to detect any trends was low due to the small sample size.

  \subsection*{Additional file 2 --- Thickness of mammary gland in false killer whales in relation to age and reproductive status.}  \label{sec:adfile2}
Thickness of mammary gland in false killer whales from South Africa in relation to age and reproductive status. Open circles are used to represent animals that were not lactating (NL) and closed circles animals for those that were lactating (L). We fitted a linear regression model to mammary gland thickness with age and reproductive class (lactating, non-lactating) as explanatory variables. The dashed horizontal line is the fitted mean mammary gland thickness in non-lactating animals, and the solid line is the fitted mean thickness for lactating animals. The grey bands around each fitted mean are the 95\% confidence intervals for the estimate. The figure shows that there is some evidence for greater mammary gland thickness in lactating animals but there was no evidence for a change with age in either.

  \subsection*{Additional file 3 --- The number of corpora lutea that represent pregnancies (CLP) and ovulation (CLO) as a function of age in false killer whales.} \label{sec:adfile3}
There were only 13 individuals in each age group (total $n$ = 26) from the combined dataset from Japan and South Africa so it is not possible to say anything conclusive about the trend in the corpora lutea of pregnancy and ovulation as function of age.

  \subsection*{Additional file 4 --- Life table for \textit{Pseudorca crassidens}} \label{sec:adfile4}
The life table for the combined dataset for specimens from Japan and South Africa with 1 year wide age classes, based on the modelled age frequency distribution. $x$ is the upper limit of the age interval, e.g., 7-8 years of age appears as 8. $fx$ is the age frequency, the number of animals of each age class in the original data, $Fx$ denotes the total number of animals at least $x$ years of age, $f'x$ denotes the fitted values from the model for the number of animals in each age class, $F'x$ denotes the total number of animals aged $x$ or older, based on $f'x$, $lx$ is the survivorship of animals of aged $x$, $dx$ is the frequency of mortality of animals aged $x$, $px$ is the age-specific survival rate, $qx$ is the age-specific mortality rate, $Lx$ is the number of individual-years lived between the ages of $x$ and $x+1$, $ex$ is the age-specific life expectancy, the number of years an individual aged $x$ is expected to still live, $Pr$ is the number of confirmed postreproductive animals.

  \subsection*{Additional file 5 --- Fitted data from smooth regression models for age-specific fecundity.} \label{sec:adfile5}
 Age-specific fecundity data had to be smoothed to generate values for each single-year age class in the survival dataset to generate the life table and carry out further analyses. It was not clear what value should be used for the degrees of freedom, in other words how smooth the plots should be, so we generated curves under 10 different scenarios for the degrees of freedom; a: df=2.1, b: df=2.9, c: df=3.7, d: df=4.5, e: df=5.3, f: df=6.1, g: df=6.9, h: df=7.7, i: df=8.5, j: df=8.9. The plot on the left shows the curves for the combined dataset and the plots on the right show the curves for the separate datasets.

  \subsection*{Additional file 6 --- Dataset of fitted values of age-specific fecundity under 10 different smoothing scenarios for the fecundity data from Japan} \label{sec:adfile6}
  \begin{verbatim} 
  AF6_Japan_fecundity_models.txt 
  \end{verbatim} 

  \subsection*{Additional file 7 --- Dataset of fitted values of age-specific fecundity under 10 different smoothing scenarios for the fecundity data from South Africa} \label{sec:adfile7}
    \begin{verbatim} 
    AF7_SouthAfrica_fecundity_models.txt 
  \end{verbatim} 
  
  \subsection*{Additional file 8 --- Dataset of fitted values of age-specific fecundity under 10 different smoothing scenarios for the fecundity data from the combined data from the two populations} \label{sec:adfile8}
   \begin{verbatim} 
   AF8_Combinedpopulation_fecundity_models.txt 
   \end{verbatim} 
  
  \subsection*{Additional file 9 --- The estimated PrR and the distribution of PrR under the null hypothesis of no PRLS for false killer whales from Japan and South Africa} \label{sec:adfile9}
The distribution of postreproductive representation, PrR, under the null hypothesis of no postreproductive life span, PRLS, is shown with a black line in each plot, with the 99\% confidence interval shaded grey, and the estimated PrR marked with a red dashed line. Each plot represents the calculation of PrR using smoothed fecundity with ten different values for the degrees of freedom in the model (i.e,. the smoothness); a: df=2.1, b: df=2.9, c: df=3.7, d: df=4.5, e: df=5.3, f: df=6.1, g: df=6.9, h: df=7.7, i: df=8.5, j: df=8.9. 

\end{spacing}
\end{document}